\documentclass[twocolumn,preprintnumbers,amsmath,amssymb]{revtex4}

\usepackage{graphicx}
\usepackage{dcolumn}
\usepackage{bm,epsfig}

\begin{document}

\title{The role of dimensionality in the Kondo Ce$TX_{2}$ family: the case of CeCd$_{1-\delta}$Sb$_{2}$}

\author{P. F. S. Rosa$^{1,2}$, R. J. Bourg$^{1}$,  C. B. R. Jesus$^{2}$, P. G. Pagliuso$^{2}$, and Z. Fisk$^{1}$}

\affiliation{
$^{1}$University of California, Irvine, California 92697-4574, U.S.A. \\
$^{2}$Instituto de F\'isica \lq\lq Gleb Wataghin\rq\rq, UNICAMP, Campinas-SP, 13083-859, Brazil.}
\date{\today}

\begin{abstract}

Motivated by the presence of competing magnetic interactions in the heavy fermion family Ce$TX_2$ ($T$ = transition metal, $X$ = pnictogen), here we study 
 the novel parent compound CeCd$_{1-\delta}$Sb$_{2}$ by combining magnetization, electrical resistivity, and heat-capacity measurements. Contrary to the antiferromagnetic (AFM) ground state observed in most members of this family, the magnetic properties of our CeCd$_{1-\delta}$Sb$_{2}$ single crystals revealed a ferromagnetic (FM) ordering at $T_{\rm c}$ =  3 K  with an unusual soft behavior.  By using a mean field model including anisotropic nearest-neighbors interactions and the tetragonal crystalline electric field  (CEF) Hamiltonian, a systematic analysis of our macroscopic data was obtained. Our fits allowed us
to extract a simple but very distinct CEF scheme, as compared to the AFM counterparts. As in the previously studied ferromagnet CeAgSb$_{2}$, a pure $|\pm 1/2 \rangle$ ground state is realized,
 hinting at a general trend within the ferromagnetic members.  We propose a general scenario  for the understanding of the magnetism in this family of compounds based on the subtle changes of dimensionality in the crystal structure.
 \end{abstract}

\maketitle

\section{INTRODUCTION}

A rich variety of ground states emerges in heavy fermion compounds due to the competition between Ruderman-Kittel-Kasuya-Yosida (RKKY) magnetic interactions, on-site Kondo interactions and crystalline electrical field (CEF) effects \cite{BookReview}. In particular, tetragonal Ce-based compounds host numerous interesting phenomena as a result of such interplay. For instance, unconventional superconductivity is observed in CeCu$_{2}$Si$_{2}$ and CeRhIn$_{5}$, quantum criticality in CeCoIn$_{5}$ and complex
antiferromagnetism (AFM) with multiple field-induced transitions in CeAgBi$_{2}$ and CeAuSb$_{2}$
\cite{CeCu2Si2,Hegger,CeCoIn5,QCP,Review115,SMThomas,Balicas}. The distance between Ce moments, i.e. the lattice parameter \textit{a}, commonly found in these compounds ranges from 4 to 5 \AA. Since RKKY interactions depend on  \textit{a}, AFM ground states and/or fluctuations are more likely to be realized in these systems, including all the examples cited above.

Nevertheless, RKKY interactions are also oscillatory in $2k_{F}a$, where $k_{F}$ is the radius of the conduction electron Fermi surface. Thus, although rare, ferromagnetic (FM) ground states have been observed in the tetragonal Kondo compounds CeRu$_{2}$Ge$_{2}$, CeAgSb$_{2}$, CeZn$_{1-\delta}$Sb$_{2}$ and, more recently, CeRuPO and CePd$_{2}$P$_{2}$ \cite{CeRu2Ge2,CeAgSb2,CeZnSb2,CeRuPO,CePd2P2}. In particular, intense efforts have been made to understand the unusual FM properties observed in CeAgSb$_{2}$, namely: (i) larger magnetic susceptibility perpendicular to [001] ($\chi_{ab}$)  despite that the magnetic ordered
moment below $T_{c}$ is parallel to the c-axis; (ii)  linear increase of the hard-axis magnetization with magnetic field below $T_{c}$ reaching $\sim 1.2$ $\mu_{B}$ at 3 T, a value much larger than the spontaneous moment along the c-axis, $0.4$ $\mu_{B}$ \cite{CeAgSb2B,CeAgSb2C,CeAgSb2D}. The origin of such intriguing properties has been elucidated by a combination of neutron scattering experiments and fits of $\chi$ to a Hamiltonian containing both CEF and anisotropic interactions terms \cite{CeAgSb2_Onuki}. The positive value of the CEF parameter B$_{2}^{0}$ accounts for $\chi_{ab} > \chi_{c}$ and an exchange interaction with strong Ising character ($J_{z} \gg J_{x,y}$) causes the magnetic ordering of the $z$ component of the angular momentum
to take over the in-plane ordering.
Moreover, an unexpected pure $| \pm 1/2 \rangle$ ground state has been shown to be realized. Whether such an unusual CEF scheme is particular to CeAgSb$_{2}$ or a general trend of ferromagnetic members in the 112 system is still an open question.

In this context, we revisit the Ce$T$X$_2$ family of compounds ($T$ = transition metal, $X$ = pnictogen) by studying and modeling the macroscopic properties of a novel member with $T=$ Cd and $X=$ Sb. As expected, CeCd$_{1-\delta}$Sb$_{2}$ is an intermetallic compound which crystallizes in the tetragonal ZrCuSi$_2$-type structure ($P4/nmm$ space group) with a stacking arrangement of CeSb-$T$-CeSb-Sb layers. As a result of its ferromagnetic order at $T_{c} = 3.0$ K, CeCd$_{1-\delta}$Sb$_{2}$ turns out to be an interesting avenue to study the evolution of FM in this family since the distance between Ce ions ($a = 4.376$ \AA) is very close to the one found in CeAgSb$_{2}$ ($a=4.363$ \AA). Moreover, by taking into account several recent reports on the Ce$T$Bi$_{2}$ family of antiferromagnetic compounds, in this work we are able to compare the results on Ce$T$Sb$_{2}$ members to previous results on Ce$T$Bi$_{2}$ compounds \cite{CeTBi2,CeNiBi2_Takabatake,Mizoguchi_CeNiBi2,Lin_RNiBi2,ReCuBi2_Camilo,Cris2014,Cris2015,RosaNi}.

To this end, here we report the physical properties of CeCd$_{1-\delta}$Sb$_{2}$ ($\delta$ = 0.3) by means of magnetic susceptibility, electrical resistivity, and specific heat measurements. Our results reveal a soft ferromagnetic ordering at $T_{\rm c}$ = 3 K with large anisotropy ratio $\chi_{ab}/ \chi_{c} = 15$ at $T_{c}$. A systematic analysis of the magnetization and specific heat data within the framework of mean field theory with the contribution of anisotropic first-neighbor interactions and tetragonal CEF allows us to extract the CEF scheme for CeCd$_{1-\delta}$Sb$_{2}$, as well as to estimate the values of the anisotropic RKKY exchange parameters between the Ce$^{3+}$ ions. Interestingly, 
 the CEF scheme obtained displays a pure $| \pm 1/2 \rangle$ ground state, as in CeAgSb$_{2}$, 
 suggesting a trend in the FM members. Our results also point out to a more general scenario where the dimensionality of the system, given by the ratio $c/a$, induces a crossover from AFM to FM order accompanied by a drastic change of ground state.

\section{EXPERIMENTAL DETAILS}

Single crystals of CeCdSb$_{2-y}$ were grown from a combined Cd/Bi-flux. The crystallographic structure was verified by X-ray powder diffraction and the extracted lattice parameters are $a = 4.376(3) $\rm{\AA} and $c = 10.903(5)  $\AA. Although the value of $a$ is very similar to the one obtained for CeAgSb$_{2}$ ($a =  4.363(1)$ \AA and $c = 10.699(5)$ \AA), the $c$ parameter is 2\% larger, likely caused by the lattice expansion due to larger transition metal ion. In addition, several samples  were submitted to elemental analysis using a commercial Energy Dispersive Spectroscopy (EDS) microprobe, which revealed the stoichiometry to be 1:0.7:2 with an error of 5\%. We note that deficiency at the transition metal site is a common trend in this family of compounds, as observed in CeZn$_{1-\delta}$Sb$_{2}$, CeAu$_{1-\delta}$Bi$_{2}$, CeNi$_{1-\delta}$Bi$_{2}$, and CeAu$_{1-\delta}$Sb$_{2}$, to name a few \cite{CeZnSb2,Cris2015,RosaNi,Seo}. Nevertheless, in CeCd$_{1-\delta}$Sb$_{2}$ the magnetic transition observed at $T_{c}$ is very sharp and both residual resistivity ($\rho_{0} = 0.4 $ $\mu\Omega$cm) and residual resistance ratio (RRR $= 76$) are consistent with a good metal.

Magnetization measurements were performed using a commercial superconducting quantum interference device (SQUID). The specific heat was measured using a commercial small mass calorimeter that employs a quasi-adiabatic thermal relaxation technique. The in-plane electrical resistivity was obtained using a low-frequency ac resistance bridge and a four-contact configuration.

\section{RESULTS AND DISCUSSIONS}

\begin{figure}
\begin{center}
\includegraphics[width=0.9\columnwidth]{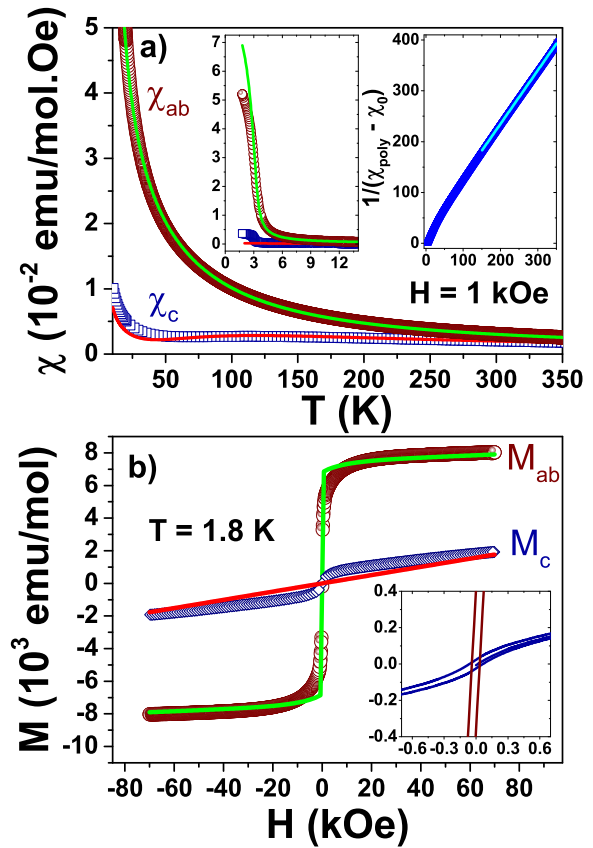}
\vspace{-0.7cm}
\end{center}
\caption{Temperature dependence of the magnetic susceptibility
measured with $H$ = 1 kOe applied parallel to the ab-plane,
$\chi_{ab}$, and parallel to the $c$-axis,  $\chi_{c}$. The inverse of the polycrystalline average 1/$\chi_{poly}(T)$ is shown in the right inset. The left inset displays the low-T behavior of $\chi$. (b) Magnetization as a function of the applied magnetic field perpendicular (open spheres) and parallel (open squares) to the $c$-axis at $T$ = 1.8 K. The inset shows the low field hysteresis. The solid lines through the experimental points are best fits of the data using the CEF mean field model discussed in the text.} 
\label{fig:Fig1}
\end{figure}

Figure ~\ref{fig:Fig1}a shows the high temperature dependence of the magnetic susceptibility, $\chi(T)$, when a magnetic field of 1 kOe is applied parallel to the c-axis, $\chi_{c}$, and along the ab-plane, $\chi_{ab}$. As in CeAgSb$_{2}$, our data show that $\chi_{ab}>\chi_{c}$, pointing out to an easy-axis in the ab-plane. It is worth noting that the opposite behavior (i.e., $\chi_{ab}<\chi_{c}$) is found in the family Ce$T$Bi$_{2}$ and in the members $T =$ Au, Ni, and Cu of Ce$T$Sb$_{2}$. Although CeAgSb$_{2}$ displays a Curie-like behavior at high temperatures for both directions, $\chi_{c}$ data of CeCdSb$_{2}$ displays an anomaly at $\sim 125$ K related to CEF effects, as we will discuss below. The left inset of Fig. 1a displays the low temperature $\chi(T)$ data, which show a clear ferromagnetic (FM) transition at $T_{\rm c}\simeq$ 3 K and a magnetic anisotropy consistent with an easy axis along the $ab$-plane. The ratio $\chi_{ab}/\chi_{c} \approx$ 15 at $T_{\rm c}$ is mainly determined by the tetragonal CEF splitting and reflects the low-$T$ Ce$^{3+}$ single ion anisotropy. The inverse of the polycrystalline 1/$\chi_{poly}(T)$ is presented in the right inset of Fig.~\ref{fig:Fig1}a. A Curie-Weiss fit to this averaged data for $T>$ 150 K (dashed line) yields an effective magnetic moment $\mu_{eff}$ = 2.5(1) $\mu_{B}$ (in agreement with the theoretical value of $\mu_{eff}$ = 2.54 $\mu_{B}$ for Ce$^{3+}$) and a paramagnetic Curie-Weiss temperature $\theta_{p}$ = -24K, which is unexpected due to the ferromagnetic order at low temperatures. We note that this value of $\theta_{p}$ is the average value found along the antiferromagnetic series 
Ce$T$Bi$_{2}$.  In a molecular field approximation, $\theta_{p}$ is proportional to the effective exchange interaction when CEF effects are averaged out. Thus, a similar value of $\theta_{p}$ may indicate similar effective exchange interactions at high temperatures, even though the magnetic ordered state is different.
We also note that negative $\theta$ has been observed in polycrystalline samples of CeZnSb$_{2}$, which recently has been shown to order ferromagnetically at $T_{c} = 3.6$ K \cite{CeZnSb2, Sologub}.

Figure 1b displays the magnetization at 1.8 K as a function of magnetic fields applied parallel to the c-axis, $M_{c}
$, and along the ab-plane, $M_{ab}$. The large magnetic anisotropy of CeCd$_{1-\delta}$Sb$_{2}$ is also evident 
in these data where $M_{ab} \gg M_{c}$. In particular, the saturation value for $M_{ab}$ reaches $1.44$ $\mu_{B}
$ while the saturation value for $M_{c}$ reaches only 0.34 $\mu_{B}$, likely to due crystal field effects and the fact that the magnetization in far from saturation, respectively. The low field magnetization, shown in 
the inset of Fig. 1b, displays an unusual soft magnetism with coercive fields of $H_{c} = 60(9)$ Oe along the c-axis 
and $H_{c} = 36(3)$ Oe along the ab-plane. The solid lines through the data points in Figs. 1 and 3 represent the 
best fits using a CEF mean field model discussed below. 

The in-plane electrical resistivity, $\rho_{ab}(T)$, of CeCd$_{1-\delta}$Sb$_{2}$ as a function of temperature is shown in Fig. 2. At high temperatures ($T >  150$ K),  $\rho(T)$ decreases linearly with decreasing temperature, as expected for metallic systems. As $T$ is further decreased, a broad feature emerges centered at $\sim 125$ K due to the thermal depopulation of the first excited CEF level. At $T_{c} = 3$ K,  $\rho_{ab}(T)$ 
 drops sharply as the magnetic scattering becomes coherent. A second kink at $T = 0.6$~K is observed in the low temperature data, as shown in the right inset of Fig. 2, which is likely due to a change in magnetic structure. In fact, it has been shown for CeZn$_{1-\delta}$Sb$_{2}$ ($T_{c} = 3.6$ K)
that the system undergoes a subsequent AFM transition at $T_{N} = 0.8$ K \cite{CeZnSb2}.

Below $T_{c}$ and above $T_{N}$, the electrical resistivity
 of CeCd$_{1-\delta}$Sb$_{2}$ can be fit by the expression:
 
 \begin{equation}
\rho(T)=\rho_{0}+ AT^{2} + D\frac{T}{\Delta}\Big(1+\frac{2T}{\Delta}\Big)e^{-\Delta/T}.
\end{equation}

The first two terms describe the usual Fermi-liquid (FL) expression. The third term is the contribution from an energy gap in the magnon dispersion relation where $D$ is related to the electron-magnon and spin disorder scattering and $\Delta$ is the magnitude of the gap.

The best fit of the data to Eq. (1) (solid lines in the left inset of Fig. 2) yields a small residual resistivity of $0.4$ $\mu\Omega$cm which, in addition to the relatively high residual resistivity ratio 
(RRR  $\equiv (\rho_{300 \mathrm{K}} - \rho_{0.5 \mathrm{K}})/\rho_{0.5 \mathrm{K}}$) of 76, indicates good crystallinity and homogeneity despite the presence of Cd deficiency. 
Interestingly, both coefficient $D$ and magnon gap $\Delta$ tend to zero, suggesting that the magnon 
contribution to the scattering is negligible. Hence, the dominant FL term yields $A = 0.1$ $\mu\Omega$cmK$^{-2}$, which is very close to the coefficient $A = 0.07$ $\mu\Omega$cmK$^{-2}$ found in CeAgSb$_{2}$. Using the reasonable assumption that the Kadowaki-Woods relation ($A/\gamma^{2} = 1.7\times 10^{-5}$ $\mu\Omega $cm(mol K/mJ)$^{2}$) is valid for CeCd$_{1-\delta}$Sb$_{2}$, one obtains $\gamma = 77$ mJ/mol 
K$^{2}$, which is only moderately heavy.

\begin{figure}
\begin{center}
\includegraphics[width=0.9\columnwidth,keepaspectratio]{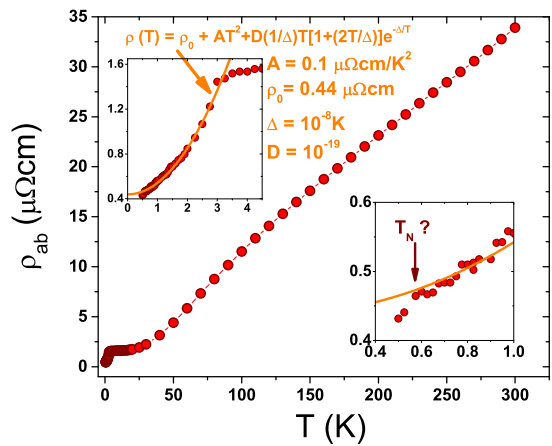}
\vspace{-0.75cm}
\end{center}
\caption{Temperature dependence of the in-plane electrical resistivity, $\rho_{ab}$. The left inset shows a fit to Eq. (1). The right inset displays the low temperature data.}
\label{fig:Fig3}
\end{figure}

\begin{figure}
\begin{center}
\includegraphics[width=0.8\columnwidth,keepaspectratio]{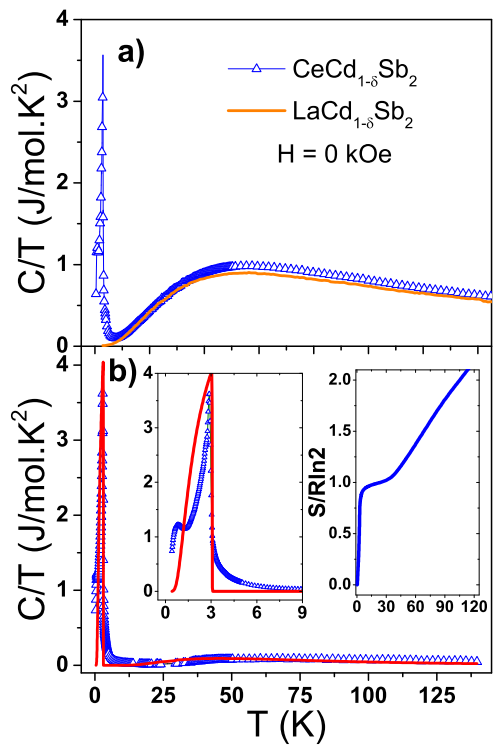}
\vspace{-0.7cm}
\end{center}
\caption{(a) $C(T)/T$ of CeCd$_{1-\delta}$Sb$_{2}$ and LaCd$_{1-\delta}$Sb$_{2}$ as a function of temperature.  (b) $C_{mag}(T)/T$ vs $T$ and the corresponding fit (solid lines) using the CEF mean field model discussed in the text. The right inset shows the magnetic entropy and the left inset the low temperature data.}
\label{fig:Fig4}
\end{figure}

In order to further explore the electronic contribution and the magnetic entropy associated with such moderate heavy fermion system, we now turn our attention to specific heat measurements. Fig. 3a shows the total specific heat divided by temperature $C(T)/T$ as a function of temperature for CeCd$_{1-\delta}$Sb$_{2}$ (open triangles) and its non-magnetic counterpart LaCd$_{1-\delta}$Sb$_{2}$ (solid lines). Fig. 3b presents the magnetic specific heat $C_{mag}(T)/T$ of CeCd$_{1-\delta}$Sb$_{2}$ after the subtraction of the lattice contribution of the La member.  In the left inset of Fig. 3b, the clear peak of $C(T)/T$ defines $T_{\rm c}$ = 3 K consistently with the both magnetization and electrical resistivity data.  The right inset of Fig. 3b shows the magnetic entropy ($S_{\mathrm{mag}}$) obtained by integrating $C_{mag}(T)/T$ over temperature. At $T_{c}$, $S_{\mathrm{mag}} \sim 80$\% of $R$ln$2$ and the entropy of the doublet is fully recovered at $\sim 7$ K. This reduction of $S_{\mathrm{mag}}$ at $T_{c}$ may be due to two indistinguishable contributions, namely, a partial compensation
of the Ce$^{3+}$ CEF ground state due to the Kondo effect and magnetic frustration/short-range interactions due to competing magnetic interactions. In fact, a second transition occurs in $C_{mag}(T)/T$ at $0.6$~K, signaling for a change in magnetic structure. At high $T$, $C_{mag}(T)/T$ displays a broad feature due to the Schottky-type anomaly resulted from the CEF splitting of $125 K$, as discussed below. In fact, $S_{\mathrm{mag}}$ at 125 K is $R$ln4, which includes the ground state and the first excited doublets.  


 \begin{table*}[!ht]
  \centering
  \begin{tabular}{c@{\hskip 0.2in}c@{\hskip 0.2in}c@{\hskip 0.2in}c@{\hskip 0.2in}c@{\hskip 0.2in}c@{\hskip 0.2in}c@{\hskip 0.2in}c@{\hskip 0.2in}}
\hline \\
\multicolumn{8}{c}{CEF parameters (in Kelvin)} \\
\cline{1-8}\\
Compound  & B$^{0}_{2}$ & B$^{0}_{4}$ & B$^{4}_{4}$ & & & $z_{FM}J_{AFM}$ & $z_{FM}J_{FM}$ \\
&&&&&&&\\
 CeCd$_{1-\delta}$Sb$_{2}$ &   12.3   &  -0.28   &  2.19  & &  & 0.2   &  -1.1        \tabularnewline
  CeAgSb$_{2}$ &   7.6   & -0.06    & $\pm$ 0.7   &  &  & -4  & -47         \tabularnewline
  CeAu$_{1-\delta}$Bi$_{1.6}$   &   -15.6   &  0.01   &  0.76  & &  & 1.4   &  -1.1        \tabularnewline
  CeCuBi$_{2}$                       &  -7.67     &  0.18  &  0.11   &  &  & 1.1  & -1.2      \tabularnewline
\hline 
\end{tabular}
  \caption{Comparison between the extracted parameters (in Kelvin) for CeCd$_{1-\delta}$Sb$_{2}$ (this work), CeAgSb$_{2}$ \cite{CeAgSb2_Onuki}, CeCuBi$_{2}$ \cite{Cris2014}, and CeAu$_{0.92}$Bi$_{1.6}$ \cite{Cris2015}. Here, $z_{AFM}$ ($z_{FM}$) are the Ce$^{3+}$ nearest neighbors with an AFM (FM) coupling.}
  \label{tab:1}
\end{table*}

 \begin{table*}[!ht]
  \centering
  \begin{tabular}{ccccccc}
\hline \\
\multicolumn{7}{c}{Energy levels and wave functions} \\
\cline{1-7}\\
$E (K)$  & $|-5/2\rangle$ & $|-3/2\rangle$ & $|-1/2\rangle$ & $|+1/2\rangle$ & $|+3/2\rangle$ & $|+5/2\rangle$ \\
&&&&&&\\
270  & -0.88   &   0.0  &   0.0 &  0.0 &    -0.47 &   0.0 \\
270  & 0.0  &  0.47 &  0.0 &  0.0 &     0.0  &   0.88 \\
128  & 0.0 & 0.88 & 0.0 & 0.0 & 0.0 & -0.47 \\
128  & -0.47  &  0.0  &  0.0 &   0.0  & 0.88 &  0.0 \\
0     & 0.0  &   0.0  &  1.0 &  0.0 &     0.0  &   0.0 \\
0     & 0.0   &   0.0  &   0.0 &  1.0 &     0.0 &   0.0 \\
\end{tabular}
  \caption{Energy level and wave functions of the CEF scheme obtained from the thermodynamic properties of CeCd$_{1-\delta}$Sb$_{2}$.}
  \label{tab:1}
\end{table*}

 In order to establish a plausible scenario for the magnetic properties of CeCd$_{1-\delta}$Sb$_{2}$, we now analyze the data presented in Figs. 1 and 3 using a mean field model including two anisotropic interactions between nearest-neighbors as well as the tetragonal CEF given by:

\begin{equation}
H_{CEF}=B_{2}^{0}O_{2}^{0} + B_{4}^{0}O_{4}^{0} + B_{4}^{4}O_{4}^{4},
\end{equation}

\noindent where $B_{i}^{n}$ are  the CEF parameters, and $O_{i}^{n}$ are the Stevens equivalent operators obtained from the angular momentum operators \cite{Stevens1}. 
 For instance, the operator $O_{2,i}^{0} = 3\hat{J}_{z,i}^{2} - J(J+1)$ favors in-plane alignment of spins
 (i.e., $\hat{J}_{z} = 0$) if $B_{20}>0$. Analogously, if $B_{20}<0$ there is a tendency of alignment
 along the $c$-axis. A more detailed description of the model can be found in Ref. \cite{Pagliuso_JAP2006}.
  
 This model was used to simultaneously fit $\chi(T)$, $M(H)$ and $C_{mag}(T)/T$ data in the entire range of temperature. The best fits yield the CEF parameters and exchange interactions displayed in Table 1, which are compared to previous analyses on CeAgSb$_{2}$, CeAu$_{1-\delta}$Bi$_{1.6}$, and 
CeCuBi$_{2}$ \cite{CeAgSb2_Onuki,Cris2014,Cris2015}. The first clear difference between these compounds is the sign of $B_{20}$, which is
negative for AFM members and positive for FM members. According to the discussion above, this result
indicates that in-plane alignment of spin is favored in FM compounds while c-axis alignment is favored
in AFM compounds. In fact, it is known from previous X-ray magnetic resonant scattering measurements on CeCuBi$_{2}$ that its magnetic structure contains spins aligned along the c-axis. Moreover, the resolved magnetic structure shows a pattern (++--) with in-plane ferromagnetic interactions and out-of-plane
antiferromagnetic interactions.
In the ferromagnetic CeAgSb$_{2}$, the $\chi_{ab}$ at higher temperatures is indeed larger than $\chi_{c}$, However, due to the strong Ising character of the exchange interactions
($J_{z} \gg J_{x,y}$), the magnetic ordering of the z component, $\hat{J}_{z}$,
takes over the ordering of the in-plane components.

Table II displays the corresponding eigenfunctions and eigenvalues of CeCd$_{1-\delta}$Sb$_{2}$. The ground state is a pure $\Gamma_{6} = |\pm 1/2\rangle$ doublet, exactly as in CeAgSb$_{2}$, followed by the first excited doublet $\Gamma^{(2)}_{7}$ $(-0.47|\pm5/2\rangle + 0.88|\mp3/2\rangle)$ at $127$~K, and the second excited doublet $\Gamma^{(1)}_{7}$  $(0.62|\pm5/2\rangle + 0.78|\mp3/2\rangle)$ at $270$~K. The obtained CEF scheme and exchange constants, $z_{AFM}*J_{AFM} = 0.2$~K  and $z_{FM}*J_{FM} = -1.1$ K~, describe well the main features of the thermodynamic data shown in Figs. 1 and 3: the value of $T_{c}$, the magnetic anisotropy of $\chi(T)$, and the Schottky anomaly in $C_{mag}(T)/T$. However, it is important to notice that the CEF parameters obtained from fits to macroscopic measurements may not be as precise and unique. An accurate determination of the CEF scheme and its parameters does require a direct measurement by, for instance, inelastic neutron scattering \cite{Christianson_CEF115}, while the mixed parameters of the wave functions may be compared with a X-Ray absorption study \cite{Severing_Ce115}.

Nonetheless, the analysis presented here clearly shows a trend in this family of compounds, as summarized in Fig. 4. In FM members, their larger values of $c$ imply that the inter-plane AFM exchange becomes unfavorable due to the larger spacing between layers. In turn, the magnetic anisotropy is inverted ($\chi_{ab} > \chi_{c}$) due to the change of sign of $B_{20}$. Finally, the ground state drastically changes from mainly $|\pm5/2\rangle$ to pure $|\pm1/2\rangle$.

More generally, the ground state depends on the interplay between both lattice parameters, i.e., on the lattice parameter ratio $c/a$, which takes into account the dimensionality of the system. As one can observed in Fig. 4, for $c/a$ values ranging from 2.1 to 2.37, AFM order is realized with ground state $|\pm5/2\rangle$ and negative $B_{20}$ values. For $c/a > 2.37$, FM ordering is favored with negative $B_{20}$ values and a pure $|\pm1/2\rangle$ ground state.

\begin{figure}[!ht]
\begin{center}
\includegraphics[width=1.\columnwidth,keepaspectratio]{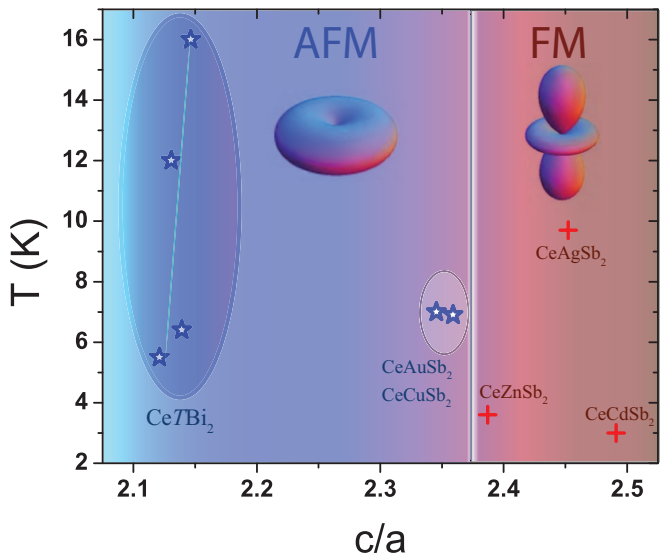}
\vspace{-0.9cm}
\end{center}
\vspace{0.5cm}
\caption{Magnetic ordering temperature $vs$ lattice parameter ratio $c/a$ for several members of the family 
Ce$TX_{2}$.}
\label{fig:Fig4}
\end{figure}

All the above arguments corroborate to the claim that the Ce$TX_2$ family of compounds presents strong local moment magnetism with a moderate Kondo compensation. The weak hybridization between the Ce$^{3+}$ 4$f$ ions and the conduction electrons likely explains why this system is less propitious to host heavy superconductivity, at least under ambient pressure. Our results also give a general scenario for the competing magnetic interactions commonly observed in these materials.

\section{CONCLUSIONS}

In summary, we have studied temperature dependent magnetic susceptibility, electrical resistivity, and heat-capacity on CeCd$_{1-\delta}$Sb$_{2}$ single crystals. Our data reveal that CeCd$_{1-\delta}$Sb$_{2}$ orders ferromagnetically at $T_{c}=$ 3.0 K and displays weak heavy fermion behavior. The detailed analysis of the macroscopic properties of CeCd$_{1-\delta}$Sb$_{2}$  using a mean field model with a tetragonal CEF suggests that the strongly localized Ce$^{3+}$ 4$f$ electrons are subjected to dominant CEF effects and anisotropic RKKY interactions. Our results shed light on the magnetic anisotropy and the role of dimensionality on the emergence of ferromagnetism in this family of compounds.

\begin{acknowledgments}
This work was supported by FAPESP (Grants No. 2013/17427-7), CNPq and CAPES-Brazil. 

\end{acknowledgments}

\bibliography{basename of .bib file}

\end{document}